\def\bfq {{\bf q}}

\def\bfp{{\bf p}}  

\def\bfr{{\bf r}}

\def\be{\begin{equation}}
 \def \ee{\end{equation}}
\def\bea{\begin{eqnarray}}
  \def\eea{\end{eqnarray}}
\newcommand{\eqn} {Eq.~(\ref}

\documentstyle[12pt,epsf,aps]{revtex}
\tighten

\begin{document}
\def\be{\begin{equation}}
 \def \ee{\end{equation}}   
\def\bea{\begin{eqnarray}}\def\eea{\end{eqnarray}}
\title{\hskip 10in 
{NT@UW-01-08},{NT@UW-03-036}\\ \vskip1cm
On the  relation between the 
Deuteron Form Factor at High Momentum
  Transfer and the High Energy Neutron-Proton Scattering Amplitude}

\author{Gerald A. Miller}
                                
\address{Department of Physics\\
  University of Washington\\
  Seattle, Washington 98195-1560}
\author{Mark Strikman}
\address{Department of Physics\\
  Pennsylvania State University,  \\
  University Park, PA 16802}
\maketitle

\begin{abstract}

A non-relativistic potential-model version of the factorization assumption,
used in perturbative QCD calculations of hadronic form factors,
 is used, along with the Born approximation valid at high energies, 
 to derive a remarkably simple 
 relationship between 
the impulse approximation contribution to 
 the deuteron form factor at high
momentum transfer  and the high energy neutron-proton scattering
amplitude.
The relation states that the form factor at a given value of 
$Q^2$  is proportional to
the scattering amplitude at a specific energy and scattering angle.
  This suggests  that 
 an accurate
computation of the form factors at large $Q^2$ requires
a simultaneous description of the  phase-shifts at a related energy,
 a statement that seems reasonable regardless of any
derivation. Our form factor-scattering amplitude 
relation is shown to be accurate for some 
examples. However, if the potential consists of a strong short distance
repulsive term and a strong longer ranged attractive  term, as typically
occurs in many realistic potentials, the relation is found to be 
 accurate only for ridiculously large values of $Q$. 
More general arguments, using only the Schroedinger equation, suggest
a strong, but complicated,
 relationship between the form factor and scattering amplitude. 
Furthermore,
the use
of recently obtained soft potentials, along with an appropriate current
operator, may allow calculations of form factors that are consistent
with the necessary phase shifts.

\end{abstract}
\newpage
\section{Introduction}
The deuteron form factor $A(Q^2)$ has been measured at 
Jefferson Laboratory and four momentum transfers
up to $Q^2=6$ GeV$^2$ \cite{Alexa:1999fe},
improving  the SLAC measurements\cite{Chertok},
 and
 measurements at larger $Q^2$ are  planned.
These efforts
have caused much 
interest  on improving calculations of the form factors
at higher values of  $Q^2$ up to about 11 GeV$^2$.
The best calculations are elegant in their use of the very latest
realistic, high-precision nucleon-nucleon potentials. 
These potentials are
based on using detailed knowledge of the long and medium range
parts of the potentials and on using artful modeling of the short distance
physics. Typically, the parameters of
the potentials are tuned to  obtain  an accurate reproduction of the
measured 
phase shifts up to 300 MeV laboratory kinetic energy, $T$.
Increasing the range of energies of the validity of the
potential should increase the ability of the potential to 
describe those aspects
of the deuteron wave function which enter at high momentum transfer\cite{limt}.

Indeed it seems reasonable to expect that describing a form factor at
a given momentum transfer, $Q^2$, would require a reproduction of
the large angle NN scattering amplitude
at 
 \begin{equation}
T \sim {Q^2\over 2 m_N},
\label{limit}
\end{equation}
where $m_N$ is the nucleon mass.
Qualitatively this is because the  relative momentum, 
which dominates
 the overlap integral for the form factor 
 is $\sim Q/2$. The implication of (1) is that for
$Q^2> 1$ GeV$^2$ one needs to use  a potential that successfully describes
the neutron-proton phase shifts for $T>500$ MeV. To our knowledge, no such
consistency check of the potential used to compute the deuteron wave function
has been made.

One might think 
that a simple kinematic relationship such as Eq.~(\ref{limit})
might not apply
because the nucleons in the deuteron are bound. 
Therefore we need  to demonstrate explicitly that there is a
strong relationship between the
form factor and the scattering 
amplitude at large energies and large 
angles. Indeed, the specific result (see Eqs.~(\ref{kin}) ~(\ref{res1})
 below)
differs quantitatively from
Eq.~(\ref{limit}).
 This is obtained  using the simplest possible
dynamics: non-relativistic spin-less nucleons interacting with an
energy-independent local potential. Clearly, it is not our
purpose  to be realistic. Instead we merely wish to point out that, under
certain conditions, the
form factor can be 
proportional to the scattering amplitude.

The derivation of the relation between the form factor and the scattering
amplitude proceeds in Sec.~II by  applying  the factorization approximation
commonly used in perturbative QCD derivations of hadronic form factors
along with the first Born approximation expected 
to be valid at high energies. The requirements for each approximation are
investigated and correction terms obtained. The
 accuracy of the approximations and the resulting form factor 
scattering amplitude relation are studied 
using simple interactions: attractive Coulomb potential and sum of attractive
and repulsive square wells in Sect.~III, and some implications for other
models are discussed. A detailed numerical study using the Malfliet-Tjon
potential is made in Sect.~IV. 
The specific approximations used in Sect.~II appear to be
 marginally successful, if the potential obtains a total weak attraction by
combining strong repulsion at
 short distances with strong attraction at larger distances. Therefore, a more
general argument, using only the Schroedinger equation,  is presented 
in Sect.~V.
A discussion of the implications of our
results is presented in Sect.~VI.

The reader may
immediately question the use of non-relativistic dynamics, because 
many specific relativistic effects are known. For a recent review, see
Ref.~\cite{Gilman:2001yh}.
 However, such
dynamics are not irrelevant at $T=500$ MeV. Furthermore,  the
use of relativistic light front dynamics shows that relativistic dynamics is
not very different from non-relativistic dynamics: two-nucleons dominate, 
there
is a wave equation, and the specific relativistic effects in the deuteron are
not very large unless $Q^2$ is very high. The specific differences between
the non-relativistic and light-front
approaches are relatively well-understood
and lead to a small  easing of the constraint (\ref{limit}) for    
$Q^2\ge m_d^2$.  Therefore,
 we turn to the necessary derivation without further apology.

\section{Basic Idea}

The deuteron wave function $\psi$ is defined by the Schroedinger equation:
\bea
\left({p^2\over 2\mu}+V\right)\psi=-\epsilon_B\psi
,
\label{psi}
\eea
where $\mu=m_N/2$.
The form factor is given in terms of the momentum-space wave function as
\bea
F(q)=\int d^3p \psi(p)\psi(\vert {\bf p+q/2}\vert)F(q)=\int d^3p 
\psi(p)\psi(\vert {\bf p+q/2}\vert)
.\label{fff}\eea
In the widely-employed Breit frame ${\bf q}^2=Q^2$.  
For large enough values of $q$,
and for a potential (expressed in momentum space)
that which decreases as a power of q,
 the integral may be simplified because
$q$ can be much larger than the typical values of $p$ for which the wave
function is near its largest value. The deuteron wave function is known to
have a limited momentum content, being essentially 0 for momenta greater
than about 600 MeV/c. 

We aim for a simplification of Eq.~(\ref{fff}) that is valid at large momentum
transfer. The basic assumption is that, for large enough values of $q$,
one can regard $q\gg p$ even though the integral extends over all values
of $p$. This leads to the factorization approximation commonly used in 
perturbative QCD
calculations of hadronic form factors.
A pedagogic
discussion of the technique we employ is presented in 
Ref.~\cite{Blp}. There are two regions for which  
the integrand of Eq.~(\ref{fff}) is largest:
$p\approx 0$ for which the first wave function is large, and
${\bf p\approx-q/2}$ for which the second wave function is large. 
The regions contribute equally so that
we may say
\bea
F(Q^2)\approx F_a(Q^2)=       2\psi(q/2)\int d^3p\;\psi(p)\label{fff1}
,\eea
with the integral over all momentum being 
proportional to  the coordinate space wave function
evaluated at the origin:
\bea
\psi(r=0)=\int {d^3p\over (2\pi)^{3\over2}}\;\psi(p).\eea

One may also  derive (\ref{fff1}) using coordinate space arguments
and determine the leading correction term. By regarding $p\ll q/2$ one finds
\cite{explain}
the approximate form factor and its leading correction term:
\bea
&&F(q)=\int d^3p \psi(p)
\int {d^3r\over (2\pi)^{3/2}}\psi(r)\exp(i{ (\bfp+\bfq/2)}\cdot\bfr))\nonumber\\
&&\approx \int d^3p \psi(p)\int {d^3r\over (2\pi)^{3/2}}\psi(r)(1-p^2r^2/6)
\exp{(i\bfq/2\cdot\bfr)}\nonumber\\
&&\approx F_a(Q^2)-(2\pi)^{3/2}\nabla^2\psi(r=0){\nabla_q^2\over6}\psi(q).\label{cap}\eea
If the integrals exist, and if 
the wave functions  fall as a power of $q$, 
 the ratio of the second to the
first terms of \eqn{cap}) is proportional to $ {\int p^2 dp\psi(p)p^2\over
\int p^2dp\psi(p)}/q^2$ which 
vanishes for sufficiently large values of $q^2$.

According to \eqn{fff1}), the momentum transfer dependence of the 
form factor is obtained
from the wave function $\psi(q/2)$, for sufficiently large values of
$Q^2$. Thus we attempt 
 to obtain $\psi(q)$ from the momentum space version
of Eq.~(\ref{psi}):
\bea
\psi(q/2)={1\over -\epsilon_B-{q^2\over 8\mu}}\int\;d^3p\;\langle {\bf q/2}\vert
  V\vert {\bf p}\rangle\;\psi(p).\label{ls}\eea
We again use the idea that  $q$ can be much greater than  $p$, so that
 the  wave function $\psi(p)$
  can be approximated as
\bea
\psi(q/2)\approx {-8\mu \over q^2} 
\langle {\bf q/2}\vert
  V\vert {\bf 0}\rangle\int d^3p\;\psi(p)={-8\mu \over q^2} 
\langle {\bf q/2}\vert
  V\vert {\bf 0}\rangle (2\pi)^{3\over2}\psi(r=0)
.\label{appr}\eea
This approximation
 seems very natural, if one is used to the ideas of perturbative QCD. 
In particular, the entire momentum transfer is taken up by  a single
action of the potential, so that the important positions in 
 coordinate space region are those for which
 the potential has its greatest variation.

It is  useful to examine \eqn{appr}) from the view of  coordinate
space. One has
\bea \int d^3p\langle {\bf q/2}\vert
  V\vert {\bf p}\rangle\;\psi(p)= \int{ d^3r\over (2\pi)^{3/2}}
\exp(-i{\bfq/2}\cdot\bfr))V(r)\psi(r).\eea
One may now observe that the
 approximation \eqn{appr}) relies on  replacing the product $V(r)\psi(r)$ by 
$V(r)\psi(0)$. This is allowed only if the potential has a much 
more significant
variation  than the wave function for positions near the origin.

The expression \eqn{appr}) 
can be used to obtain the form factor from Eq.~(\ref{fff}). The result is
\bea
F_b(Q^2)\approx \left({-16\mu\over q^2}\right)\langle {\bfq/2}\vert
  V\vert {\bf 0}\rangle\; \Phi^2,\label{res}\\
  \Phi\equiv \int d^3p\; \psi(p).\eea
The
approximation (\ref{res})
has been verified numerically
 in Ref.~\cite{tm} for the case of the
semi-realistic Hulth\'{e}n model.

The content of Eq.~(\ref{res}) is that the form factor is 
proportional to 
the square of
the bound state wave function at the origin  
times the Fourier
transform of the potential:
\bea \langle
 {\bf q/2}\vert
  V\vert {\bf 0}\rangle ={1\over (2\pi)^3}\int d^3r\; e^{{i\over 2}
{\bf q}\cdot{\bf
      r}}V(r)\\ 
={4\pi\over (2\pi)^3}\int r^2 \;dr{ \sin{{1\over2}qr}\over {1\over2} qr}
V(r).\eea

The desired relation between the form factor and the scattering
 amplitude can 
be obtained by realizing that at   energies E
much greater than the characteristic strength of the 
potential, the first Born
approximation is valid and the scattering amplitude for a cm angle $\theta$,
$f_E(\theta)=f_E({\bf k},{\bf k'}),\; E=k^2/2\mu=k'^2/2\mu,\;
{\bf k}\cdot{\bf k'} =k^2\cos\theta$
is also a Fourier transform of the potential. The validity of the first Born
approximation, at high energies, for energy-dependent local potentials,
is well-verified in many quantum mechanics textbooks. In that case, 
\bea
f_E(\theta)
\approx -{4\pi^2\mu\over (2\pi)^3}\int  d^3r  e^{i{\bf( k-k')}\cdot{\bf r}}
V(r).\label{born}
\eea            
The relation between $f_E(\theta)$ and $\langle
{\bf q/2}\vert  V\vert {\bf 0}\rangle$ 
is obtained by specifying
\bea
{1\over 4}q^2=2k^2(1-\cos\theta)=4\mu E (1-\cos\theta)=Tm_N(1-\cos\theta).
\label{kin}
\eea           
Imposing the relation (\ref{kin})
 immediately yields
\bea
f_E(\theta)=-{4\pi^2\mu}\langle{\bf q/2}\vert  V\vert {\bf 0}\rangle,
  \eea                          
and with  Eq.~(\ref{res})
\bea
F(Q^2)={1\over \pi^2q^2}f_E(\theta)\; \Phi^2.
\label{res1}
\eea
This is the result
we have been seeking. The key point is that the form factor is expressed as
 the high-energy scattering amplitude times  
well-defined factors, implying 
that 
a correct calculation for the form factor can only be achieved in models in
which the scattering amplitude is accurately reproduced. 

\section{Toy Models}

The arguments used in the previous section are analogous to those presented
in derivations of QCD factorization theorems. However, these have not
been used often for nuclear targets. We therefore discuss two simple
examples.

\subsection{Coulomb binding}
We take the potential to be
\bea
V(r)=-{g^2\over r},\eea
with the exact wave function given by
\bea
\psi(r)=N e^{-r/a},\eea
with  $a=1/(\mu g^2)$. The form factor $F(q)$ of \eqn{fff}) is given by
\bea
F(q^2)={1\over \left(a^2q^2 +{16}\right)^2}\eea
The wave function in momentum space is
\bea
\psi(p)={4\pi N\over(2\pi)^{3/2}}{2\over a\left(q^2+{1\over a^2}\right)^2}
,\eea
so that the approximation of  \eqn{fff1}) yields 
\bea
F_a(q^2)={1\over \left(a^2q^2 +{4}\right)}\eea
$F_a$ is close to $F$ for $q^2a^2\gg 3$. Using $\epsilon_B=a^2/2\mu$ and the
deuteron binding energy and $\mu=M_N/2$ means that all that is required is
$q\gg75$ MeV/c.
The approximation of \eqn{res}) yields
\bea F_b (q^2)={1\over a^4q^4},\eea
which valid under similar conditions.

The next task is to determine  the conditions that the first 
Born approximation \eqn{born}) be valid. It is well known that, 
for the Coulomb interaction, this approximation reproduces
the exact scattering cross section. However, the correct scattering 
amplitude is complex, while the approximation \eqn{born}) gives a real 
result. Thus the condition that  \eqn{born}) be valid is the condition that the
s-wave scattering phase shift be small. This is the condition that
$g^2M_N/(2k)\ll1$, or $ka\gg1$. Since $a=1/(45{\rm MeV/c}$, one  
needs only $k\gg 45 MeV/c$.

Thus the approximations of the previous section are easily satisfied for 
dynamics defined  by the Coulomb potential. More generally, it is 
reasonable to expect that, if the potential is local, purely attractive
with a significant gradient at the origin and 
magnitude determined by the very 
small deuteron binding energy (2.2 MeV), the approximations needed to
reach the form factor scattering amplitude relation (\ref{res1}) are well
satisfied. In such cases, the potential varies more rapidly than the
wave function for positions near the origin, and the potential is 
weak enough so that the first Born approximation can be accurate for
reasonable momentum transfers.

\subsection{Short-ranged repulsive square well plus long ranged attractive
square well}

The nuclear force is repulsive at short distances and attractive at long
distances between nucleons. Thus we   consider the model
defined by 
\bea V(r)=V_0 \theta(R_0-r) -V_1\theta(R_1-r)\theta(r-R_0), V_{0,1}>0.
\label{square}\eea
We take $R_0=0.4$ fm and $R_1=1.5 $ fm. Then the observed deuteron
binding energy is 
reproduced using $V_0=0.302 ~\rm{fm}^{-1},\;V_1=29.18~\rm{fm}^{-1}=5.757 $ GeV.  
This   corresponds to a hard core repulsion. The wave function $\psi(r)$
is given (up to an overall normalization constant) by 
\bea \psi(r)={\sin KR_0\over \sin KR_1}
{\sinh \gamma r\over \sinh\gamma R_0},\;r<R_0\nonumber\\
={\sin K r\over \sin K R_1},\; R_0\le r<R_1\nonumber\\
={\exp (-r/a)\over \exp(-R_1/a)},\;r\ge R_1\label{psir}\eea 
\begin{figure}
\unitlength.8cm
\begin{picture}(8,12)(-16,-0)
\includegraphics{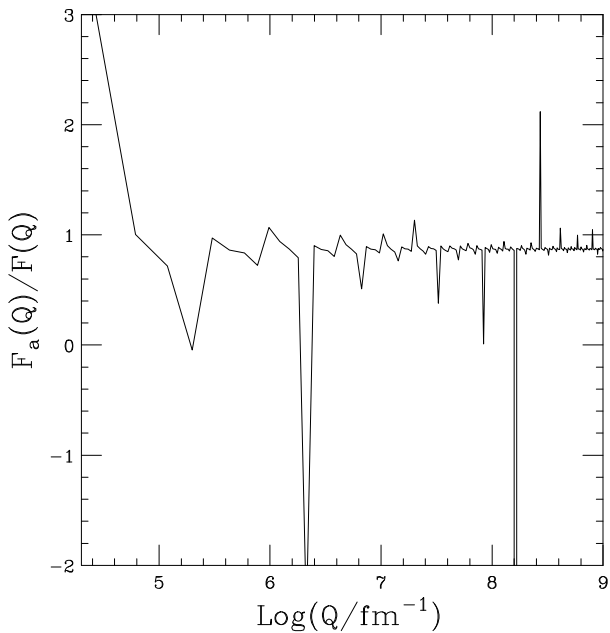}
\end{picture}
\caption{\label{figg} Ratio of approximate to exact form
factor.}
\end{figure}

The  exact ($F(q)$) and approximate ($F_a(q)$ \eqn{fff1}))
 form factors are compared by displaying their ratio
in Fig.~1. The are some wild fluctuations arising from nodes of $F(Q)$,
but these are not the most interesting feature. The true difficulty is 
that
the approximation becomes valid only 
for extremely large values of
$q> 100 $ fm$^{-1}$ .
 Furthermore, the ratio  $F_a(q)/F(q)$ is =0.87 instead of unity.
That  difficulties exist can be seen immediately by comparing
Eqs.~(\ref{fff1}) and (\ref{psir}). The coordinate-space wave function
vanishes at the origin, so the leading approximation vanishes! 
This signals a breakdown of the derivation. The point is that the 
factorization Eq.~(\ref{fff1})  depends on the potential varying more 
rapidly near the origin than the wave function

One might then be amazed that the approximate form factor is even close to
the exact one. That this occurs can be understood from the discussion below
\eqn{cap}). The action of $\nabla_q^2$ does not introduce an extra factor of
$1/Q^2$ because the momentum space wave function
contains terms $\sim {\sin QR_{0,1}\over Q^2}$.

This section contains two simple models. 
The approximations work magnificently
if  the   Coulomb potential is used, but fail miserably with
the nucleon-force-motivated 
model of \eqn{square}). Let's examine a more realistic model to see if
our approximations are relevant for 
understanding deuteron physics.

\section{Malfliet-Tjon Potential}

The short-distance repulsion and longer-ranged attraction are simulated via
potentials of the Yukawa form in the Malfliet-Tjon potential
\cite{Malfliet:1968tj}:
\bea
V(r) =-\lambda_A{ e^{-\mu_A r}\over r}+\lambda_R {e^{-\mu_R r}\over r}.\eea
The parameters were chosen to reproduce the 
deuteron binding energy, scattering length, effective range
 and
phase shifts up to a laboratory energy of 300 MeV. This potential is much
smoother than the square wells of \eqn{square}).
 
\begin{figure}
\unitlength.8cm
\begin{picture}(8,12)(-0,-13)
\includegraphics{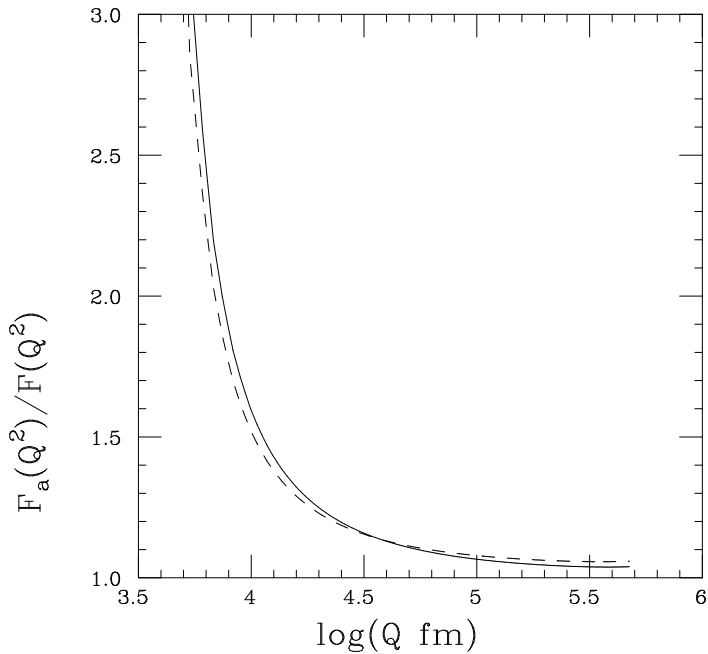}
\end{picture}
\caption{\label{figfacomp} Ratio of approximate to exact form
factor for the Malfliet-Tjon deuteron wave function. The solid curve
compares  
 the approximation of \eqn{res1}) with the exact form factor. The dashed
curve compares  
 the approximation of \eqn{fff1}) with the exact form factor.}
\end{figure}

Our aim is to study the approximations given in Eqs.~(\ref{fff1},\ref{appr}) and
the Born approximation of \eqn{born}).
 The comparison between the exact form factor $F(Q^2)$ and the
 approximation
$F_a(Q^2)$ of \eqn{fff1}) is displayed in Fig.~\ref{figfacomp}.

The approximation does work, but only for 
huge values of $Q$. The value of $Q\sim 140 $ fm$^{-1}$, or $Q^2\approx 900
$ (GeV/c)$^2$
for its natural logarithm to reach the value of $5$.
One may search for the cause of this bizarre limit by examining 
the correction
term shown of \eqn{cap}). This
 is governed by the kinetic energy operator acting 
on the wave function at the origin. The Schroedinger equation that this
essentially
 the potential times the wave function,  with a product that 
varies as $(\lambda_R-\lambda_A) $ near the origin. The strong nature of the
repulsive term $\lambda_R=7.41$ causes   the
expansion shown in  \eqn{cap})to converge very slowly.

\begin{figure}
\unitlength.8cm
\begin{picture}(8,12)(-0,-13)
\includegraphics{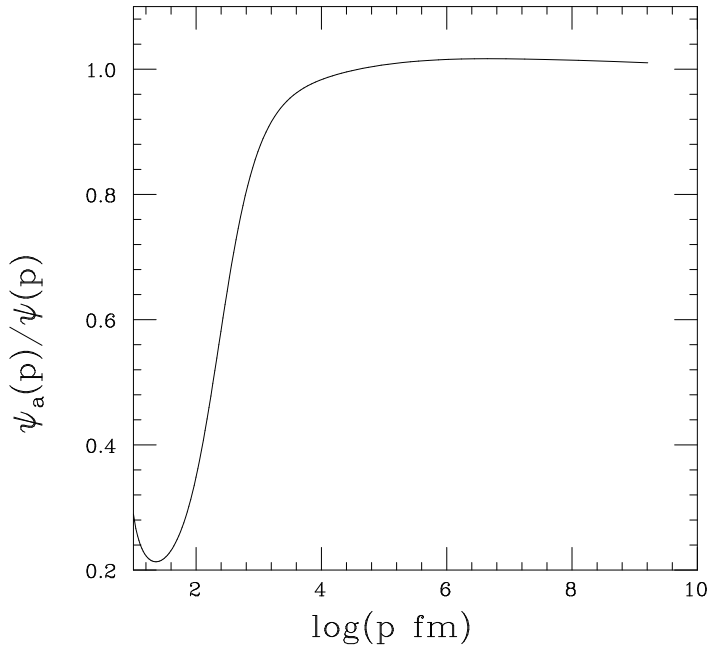}
\end{picture}
\caption{\label{psiacomp} Ratio of approximate to exact 
wave function  for the Malfliet-Tjon deuteron wave function.}
\end{figure}

One may gain further insight by studying the relationship between 
the exact and approximate wave function of \eqn{appr}). As shown in
Fig.~\ref{psiacomp}, the approximation attains validity only at supremely
 large
values of the relative momentum. Here accuracy 
 requires 
$V(r)\psi(r)\approx V(r)\psi(0)$ for values of $r$ near the origin.
For the Malfliet-Tjon potential $V(r)\sim 1/r$ and $\psi(r)$ approaches  0,
 in contrast to  the  Coulomb
wave function. From this, 
the failure of the approximation seems natural. Moreover,
many  realistic potentials behave in a similar fashion. 

\begin{figure}
\unitlength.8cm
\begin{picture}(8,12)(-0,-13)
\includegraphics{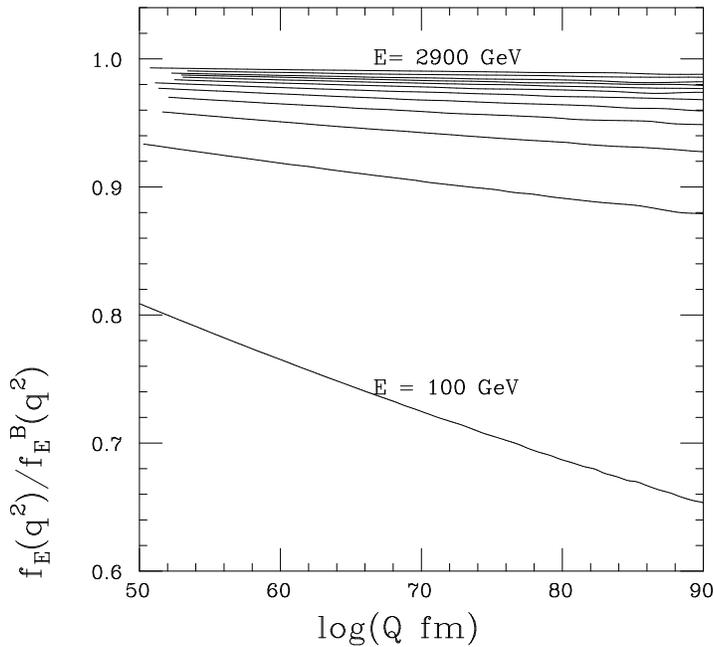}
\end{picture}
\caption{\label{Borna} Ratio of approximate to the real part of the 
 exact forward scattering amplitude
for the Malfliet-Tjon potential. E is a laboratory energy.}
\end{figure}

The final requirement needed to achieve the result (\ref{res1})
is the validity of the Born approximation. 
Our comparison of the exact and Born approximation to the forward
scattering amplitude is shown in Fig.~\ref{Borna}. 
The partial wave expansion
is used unless the   energy is
 large enough  (lab energy greater than one or two GeV) for 
the eikonal approximation to reproduce  the
exact scattering amplitude.  Fig.~\ref{Borna} shows that 
the Born 
approximation does become valid, but only at absurdly  high energies.
This figure shows results only for forward scattering, whereas our relation
(\ref{res1}) requires high momentum transfer. The convergence
to the Born approximation occurs for higher energies for large scattering
angles, if the Malfliet-Tjon potential is used.

 The slow approach to the Born approximation can be understood from the
eikonal formalism in which the scattering amplitude is expressed in terms
of an integral 
\bea
f_E(\theta)=-ik\int_0^\infty bdb\;J_0(2kb \sin(\theta/2))(e^{i\chi(b)}-1),
\eea
with the phase shift function $\chi(b)$ given as
\bea
\chi(b)=-{\mu\over k}\int_0^\infty dz V(\sqrt{b^2+z^2}).\eea
For the Malfliet-Tjon potential, $\chi(b)$  is given by 
\bea
\chi(b)={m\over k}(\lambda_R K_0(b\mu_R)-\lambda_A K_0(b \mu_A)),
\eea
which is small only for very large values of the relative momentum  $k$.
While the 
overall potential is able to produce only one weakly bound state, the
weak attraction arises  from the cancellation of two very strong terms
of opposite sign that generally  causes  large values of $\chi(b)$. 

The factorization approximations for the 
Eqs.~(\ref{fff1},\ref{appr}) and
the Born approximation of \eqn{born}) work for sufficiently large values
of the transferred momenta. A reasonable reader might examine the  figures
2-4 and conclude that ``sufficiently large'' means too large to be relevant
for real experiments. 
However,  success at asymptotically high momenta indicates
that the key assumption Eqs.~(\ref{fff1},\ref{appr}) that the 
form factor is proportional to the wave function at the origin 
 is a very
 stringent  requirement that is especially difficult to satisfy for
a wave function that vanishes at the origin. 
Furthermore, high momentum transfer scattering may   proceed
via a set of
small momentum transfer processes under certain circumstances. Thus 
the condition that the Born approximation (scattering by a single action of the
potential) be valid is also a very strong one,
as is spectacularly manifest in Fig.~4. Truly magnificently high 
 energies are required for the Born approximation to be valid.

While the relation \eqn{res1}) suggests a close connection between the form
factor and the scattering amplitude it is worthwhile to ask if one can
 derive a connection between the form
factor and the scattering amplitude 
that does not directly involve the wave function at the origin and 
single-scattering  assumptions. The desired connection is
displayed in the next section.

\section{General Approach}

The central  idea is that the wave function is dominated by its low momentum
components and that high momentum components are obtained by 
at most one  high momentum transfer operation.  We intend to manipulate the
Schroedinger equation to derive an explicit relation
 between  the form factor 
 and the scattering amplitude.
Projection operators are used to develop the necessary formalism. Let
\bea Q\equiv \int d^3p\theta (\Lambda-p) \vert \bfp\rangle\langle \bfp\vert,
\quad
P\equiv \int d^3p\theta (p-\Lambda) \vert \bfp\rangle\langle \bfp\vert,\eea
with $P$ a projection operator on high momentum states and $\Lambda$ a 
parameter, approximately 600 MeV/c or higher,  denoting
 the separation between
the high and low momentum transfer regions. Then the Schroedinger equation can
be expressed in terms of low, 
$Q\vert\psi\rangle\equiv\vert\psi\rangle_Q,$ and  high, 
$P\vert\psi\rangle\equiv\vert\psi\rangle_P,$ momentum components:
\bea
\vert\psi\rangle_P={1\over -\epsilon_B- T_{PP} }\left(V_{PP}\vert\psi\rangle_P
+V_{PQ} \vert\psi\rangle_Q\right)\label{one}\\
\vert\psi\rangle_Q={1\over -\epsilon_B- T_{QQ}}\left( V_{QQ}\vert\psi\rangle_Q
 +V_{QP} \vert\psi\rangle_P\right), \label{two}
\eea
where the notation $PTP=T_{PP}, PVP=V_{PP}$, etc. is used for the kinetic
energy $T$ and potential energy operators. 
 Substitute \eqn{two}) into \eqn{one}) to obtain:
\bea
\vert\psi\rangle_P={1\over -\epsilon_B- T_{PP}}
\left(V_{PP}+V_{PQ}
{1\over -\epsilon_B- T_{QQ}-V_{QQ}}V_{QP}\right) \vert\psi\rangle_P
.\label{threet}
\eea
Consider the complete
eigenstates of the Hermitian, energy-independent operator
$T_{QQ}+V_{QQ}:$
\bea
E_k\vert\phi_k\rangle^{(-)}=(T_{QQ}+V_{QQ})\vert\phi_k\rangle^{(-)}\\
E_n\vert\phi_n\rangle=(T_{QQ}+V_{QQ})\vert\phi_n\rangle
\eea
in which $\vert\phi_k\rangle^{(-)}$ is a scattering state with 
incoming boundary conditions, $E_k=k^2/2\mu$, and 
$E_n\ne-\epsilon_B,\vert\phi_n\rangle$ are
 the energies and wave functions of any bound states that may exist.
The use of completeness in \eqn{threet}) leads to
\bea
\vert\psi\rangle_P={1\over -\epsilon_B- T_{PP}}
\left(V_{PP}+\int d^3k
{V_{PQ}\vert\phi_k\rangle^{(-)}\;^{(-)}\langle\phi_k\vert V_{QP}
\over -\epsilon_B- E_k}+
\sum_n{V_{PQ}\vert\phi_n\rangle\;\langle\phi_n\vert V_{QP}
\over -\epsilon_B- E_n}\right) \vert\psi\rangle_P.
\label{big}\eea

The form factor depends on the high  momentum components of the deuteron wave
function, so consider the quantity $\langle \bfq/2\vert\psi\rangle_P$ for
$Q/2>\Lambda$. Then 
$\langle \bfq/2\vert\psi\rangle_P=\langle \bfq/2\vert\psi\rangle$.  
The second term on the right-hand-side of \eqn{big}) is determined by the
matrix elements $\langle \bfq/2\vert V\vert\phi_k\rangle^{(-)}$, which
are scattering amplitudes for off-energy-shell kinematics\cite{explain1}. 
These matrix elements 
are the terms which contain the relation between the form  factor and the
scattering amplitude.
 To see this, take the parameter 
$\Lambda$ to be just a bit less than $q/2$. The largest of the matrix elements
$\langle \bfq/2\vert V\vert\phi_k\rangle^{(-)}$ will be those for which
$k$ is just a bit less than $\Lambda$. In this case, the kinematics are nearly
on-shell. Transition  matrix elements are continuous, so that it is
reasonable to expect that reproducing the observed scattering amplitudes
at high energy is important for reproducing the important features of the
high momentum components of bound state wave functions.
\section{Discussion}

The use of two factorization approximations, Eqs.~(\ref{fff1}) and
(\ref{appr}) combined with the first Born approximation,\eqn{born}),
lead to a statement (\ref{res1}) that the form factor is proportional
to the scattering amplitude times the square of the coordinate space
wave function at the origin  divided by $Q^2$.
A relation very similar to Eq.~(\ref{res1}) was derived long ago 
\cite{Brodsky:1976mn} using a scale-invariant, six-quark model.
It is also true that 
the 
relation between the deuteron form factor and the scattering amplitude
has also been the subject of Ref.~\cite{aaa} in which   dispersion
relations are used
to compute the deuteron charge form factor with  experimental
phase shifts as the essential input.
Thus one sees the close relation between $F(Q^2)$ and
$f_E$ from a variety of different approaches: non-relativistic dynamics,
light-front dynamics, 
quark models and dispersion relations.

But the  validity of  Eqn.~(\ref{res1}) depends on the use 
of  factorization
approximations and the Born approximation that might not be 
valid for realistic nucleon-nucleon 
potentials. The analysis of toy models,  Sects. II, and the Malfliet-Tjon
potential Sect. III, suggests that only a  soft potential 
can be expected to satisfy the conditions necessary to obtain the 
result (\ref{res1}). Until recently, the idea  that a soft potential could
also be realistic was only a faint hope. But recent work, using effective
field theory have introduced a set of soft-realistic potentials
\cite{Bogner:2003wn},
\cite{Walzl:2000cx},
\cite{Entem:2003ft}. It is possible that these potentials, which
do not contain those features of the Malfliet-Tjon potential that
violate the necessary approximations,
could  satisfy  Eqn.~(\ref{res1}).

It is worthwhile to suppose that 
Eq.~(\ref{res1}) could be  accurate for some soft realistic potential.
Then we may
estimate the kinetic energies required for  reproducing the
phase shifts. Since we are dealing with the S-wave deuteron wave 
function we can take   $\cos \theta=0,\;$ to 
correspond to the maximal momentum transfer for identical particles. 
In this case, Eq.(\ref{kin}) gives
\be Q^2= 4m_N  T.\label{newkin}\ee
The use of this is simple. Suppose the phase shifts are well
described up to about $T=350 $ MeV, which is the standard stated upper limit.
Then one can calculate the form factor
up to only $Q^2\approx 1.4 \;{\rm GeV}^2$. But modern high-precision $NN$
potentials actually describe the data up to laboratory energies of
about 1 GeV\cite{rm,more}. 
Thus one may compute the form factors up to $Q^2\approx
4 \;{\rm GeV}^2$. 
However,  present measurements reach $Q^2=6\;{\rm GeV}^2$, and
there are plans to reach higher values. Thus the constraints we present may
 be
relevant for present and future measurements.

An additional worry, is
that the analysis of the previous paragraph might
 require a relativistic treatment. 
If we instead  were to use
light cone models as
in \cite{FS78,coester} we would find that up to 
$Q^2\sim m_d^2$ the light cone fractions of the nucleons are 
approximately equal and the relation  Eq.~(\ref{kin}) holds, but  with
$T=2E+E^2/2m_N$. 
At larger values of $Q^2$, the  increase of the effective 
invariant energy with $Q^2$ decreases somewhat. 
In any case, one can see that for $Q^2\ge 2$ GeV$^2$
one reaches the region
 where masses in the intermediate state exceed 3 GeV and
 the legitimacy of
 the two nucleon approximation becomes highly questionable, as discussed in
Ref.~\cite{FS78}. Then one would need to include either new hadronic degrees 
of freedom in the deuteron wave function: $\pi NN, \Delta\Delta,...$, 
 or to account explicitly for 
quark-gluon degrees
 of freedom.

The implications of our result (\ref{res1}) is that potentials used to obtain
the deuteron wave function should be tested by computing the corresponding
phase shifts. If one wants an accurate calculation, the phase shifts need
to be correctly obtained up to kinetic energies given by Eq.~(\ref{newkin}). 
This requirement appears, according to the arguments of
Sect.~V,  
to be more general (see \eqn{big})
 than accuracy of the asymptotic
 relation  between the scattering amplitude and the form factor, (\ref{res1})
as 
the slow onset of asymptotia occurs, in part,  because the wave function at the
origin is severely suppressed by strong short distance repulsion, and because
 single hard scattering is not strongly dominant over the possibility of
achieving a high momentum transfer as a 
result of several small momentum transfer processes.
large extent due to cancellations between
the contribution dominating in the asymptotic region and
other contribution.
All together this
suggests
that potentials currently employed to compute deuteron
form factors be tested for consistency with high energy phase shifts.
A soft potential \cite{Bogner:2003wn},
\cite{Walzl:2000cx},
\cite{Entem:2003ft} and an appropriately derived current operator should also 
be
used.

\section*{Acknowledgments}
This work is partially supported by the USDOE. 
MS thanks the INT for hospitality during the time this work was completed.
We thank Rupert Machleidt for useful discussions.


\begin{references}
\bibitem{limt} Our focus is on the nucleonic contribution to the form factors,
so we do not consider the effects of meson exchange currents.
\bibitem{explain} The notation $\langle\bfr\vert\psi\rangle=\psi(\bfr)\;
\langle\bfp\vert\psi\rangle=\psi(\bfp)$ is employed. The arguments  
$\bfq,\bfp$ refer to momentum space wave functions,
 while $\psi(\bfr)$ is a coordinate space wave function.
\bibitem{more} Such potentials are energy-dependent, caused by, for example, an
$N^*N$ intermediate state. In that case, one would need to incorporate
a correct description of the production of such a state, and 
the kinematics of Eq.~(12) would need to be revised. Thus  an estimate closer
to that of Eq.~(1) would be obtained.
\bibitem{explain1} These scattering amplitudes are not the physical ones because
the Hamiltonian is $QHQ$ instead of $H$. However, the low-momentum part of the
Hamiltonian contains the dominant interactions so that it is reasonable to
expect a close association between the scattering amplitudes of $QHQ$ and
those  of $H$.
\end{references}
  \end{document}